\documentclass[prb,showpacs,twocolumn,byrevtex]{revtex4}
\usepackage{graphicx}
\usepackage{float}
\usepackage{bm}
\usepackage{epsfig} 
\usepackage[all]{nowidow} 
\usepackage{multirow}

\begin{document}

\bibliographystyle{apsrev}


%
%

\title[Observation of cyclotron antiresonance in the topological insulator Bi$_2$Te$_3$]
{Observation of cyclotron antiresonance in the topological insulator Bi$_2$Te$_3$}

%
%
%
\author{S.V. Dordevic}
\address{Department of Physics, The University of Akron,
Akron, Ohio 44325 USA}%
\author{Hechang Lei$^{\ast}$}
\address{Condensed Matter Physics and Materials Science
Department, Brookhaven National Laboratory, Upton, New York 11973 USA}%
\author{C. Petrovic}
\address{Condensed Matter Physics and Materials Science
Department, Brookhaven National Laboratory, Upton, New York 11973 USA}%
\author{J. Ludwig}
\address{National High Magnetic Field Laboratory, Tallahassee,
Florida 32310 USA}%
\author{Z.Q. Li$^{\dag}$}
\address{National High Magnetic Field Laboratory, Tallahassee,
Florida 32310 USA}%
\author{D. Smirnov}
\address{National High Magnetic Field Laboratory, Tallahassee,
Florida 32310 USA}%

\date{\today}

%
%
\begin{abstract}
We report on the experimental observation of a cyclotron
antiresonance in a canonical 3D topological insulator
Bi$_2$Te$_3$. Magneto-reflectance response of single crystal Bi$_2$Te$_3$
was studied in 18 Tesla magnetic field, and compared to other
topological insulators studied before, the main spectral
feature is inverted. We refer to it as an antiresonance.
In order to describe this unconventional behavior we
propose the idea of an imaginary cyclotron resonance frequency,
which on the other hand indicates that the form of the Lorentz
force that magnetic field exerts on charge carriers takes
an unconventional form.
\end{abstract}

%
%
%
%
%

\maketitle

\section{Introduction}

For decades cyclotron resonance has been an important experimental tool
in plasma \cite{guest-book,hartmann-book,lominadze-book}
and condensed matter physics \cite{cardona-book,cr-book}.
In condensed matter physics for example, cyclotron resonance has been used
for probing charge dynamics, in particular in
semiconductors. Valuable information about band structure,
carrier scattering rate, effective mass, etc. can be obtained
from these measurements. In recent years, with the advent of materials
with Dirac electrons like graphene
and topological insulators, cyclotron resonance measurements have
been extensively used for probing their electronic structure.
Cyclotron resonance has been observed and characterized
in single and multiple layer graphene \cite{kuzmenko11}, thin films of
Bi$_2$Se$_3$ \cite{armitage15}, bulk Bi$_2$Se$_3$ \cite{dordevic16},
elemental bismuth \cite{laforge12,kuzmenko14,kuzmenko16},
Bi$_{1-x}$Sb$_x$ \cite{schafgans12,dordevic12b}, Bi$_{1-x}$As$_x$
\cite{dordevic14}, (Bi$_{1-x}$Sb$_x$)$_2$Te$_3$ \cite{shao17},
nano-flakes of Bi$_2$Te$_3$ \cite{tung16}, etc.

In this work we used magneto-optical spectroscopy to study
topological insulator Bi$_2$Te$_3$.
Magneto-optical activity was detected in reflection spectra,
but surprisingly
in Bi$_2$Te$_3$ we observed cyclotron {\it antiresonance},
where the spectral feature is inverted, i.e. it is a mirror image
of the resonance observed in other systems. For comparison
we also measured another canonical topological insulator
from the same family Sb$_2$Te$_3$, and we observed a
conventional resonance. Based on our data analysis
we suggest that the antiresonance in Bi$_2$Te$_3$ is due
to an unconventional form of the Lorentz force that external
magnetic field exerts on the charge carriers.

\section{Experimental results}

Single crystals of Bi$_2$Te$_3$ and Sb$_2$Te$_3$ were grown at
Brookhaven National Laboratory \cite{fisk89,canfield92} and
characterized with X-ray diffraction using a
Rigaku Miniflex X-ray machine. The analysis showed that samples
were single phase, and with lattice parameters consistent with the
previously published values \cite{hunter98}. Samples had a thickness
of several millimeters, and naturally flat surfaces with a typical
size of about 3~mm. Before every spectroscopic measurement
the samples were mechanically cleaved in order to expose a fresh surface.

Far-infrared and mid-infrared magneto-reflectance ratios
R($\omega$,B)/R($\omega$, 0 T) were collected at the National High Magnetic
Field Laboratory using superconducting 18 T magnet.
Reflectance ratios provide the most direct evidence for magneto-optical
activity, as they do not require any data analysis or manipulation. All measurements were
performed at 5 K, with unpolarized light and with the electric field vector
parallel to the quintuple layers of Bi$_2$Te$_3$ and Sb$_2$Te$_3$,
i.e. in Faraday geometry.

\begin{figure}[t]
\vspace*{-1.0cm}%
\centerline{\includegraphics[width=9.5cm]{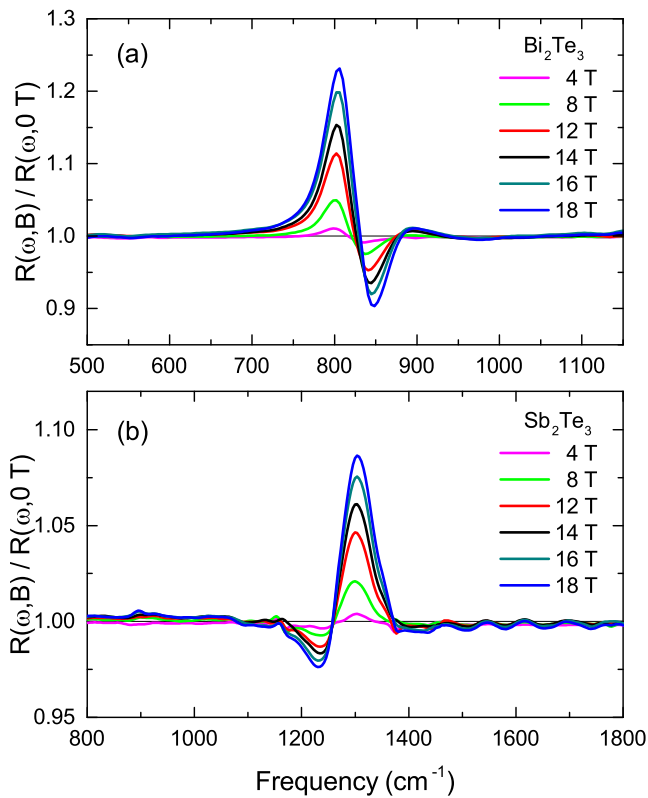}}%
\vspace*{-0.5cm}%
\caption{(Color online). (a) Reflection ratios of Bi$_2$Te$_3$ at several
magnetic field values. The feature around 800 cm$^{-1}$
is identified as cyclotron antiresonance, because it is a mirror image of
the feature observed in Sb$_2$Te$_3$ (bottom panel).
(b)  Reflection ratios of Sb$_2$Te$_3$ at several
magnetic field values. The feature around 1,300~cm$^{-1}$
is identified as the usual cyclotron resonance. Note different
axes scales between the two panels.}
\vspace*{0.0cm}%
\label{fig:bite}
\end{figure}

Figure~\ref{fig:bite} shows the infrared reflectance
ratios R($\omega$, B)/R($\omega$, 0 T) of Bi$_2$Te$_3$ and Sb$_2$Te$_3$
in magnetic fields up to 18 T. In Bi$_2$Te$_3$ we observed
magnetic field induced changes in reflectance (Fig.~\ref{fig:bite}(a)),
exceeding 20 $\%$ in 18 T,
in the region around 800 cm$^{-1}$ (100 meV). This is precisely the region
where the plasma minimum was observed in the zero-field reflectance \cite{dordevic13}.
In Fig.~\ref{fig:bite}(b) we display the ratios for Sb$_2$Te$_3$
and they also show field induced changes with the maximum change of about 10 $\%$
around 1,300 cm$^{-1}$ (160 meV), which is the location of
plasma minimum in zero field reflectance \cite{dordevic13}.
In Sb$_2$Te$_3$ conventional cyclotron resonance is
observed, similar to other systems, such as graphene,
Bi$_2$Se$_3$, bismuth, Bi$_{1-x}$Sb$_x$, etc.
The resonance in these systems manifest
as a characteristic dip-peak structure in reflectance ratios.
Contrary to all of them, in Bi$_2$Te$_3$ we observe the exactly
opposite: a peak-dip structure, and we refer to it as
an antiresonance \cite{comm-flake}.


\section{Discussion}

\begin{figure}[t]
\vspace*{-0.5cm}%
\centerline{\includegraphics[width=9.5cm]{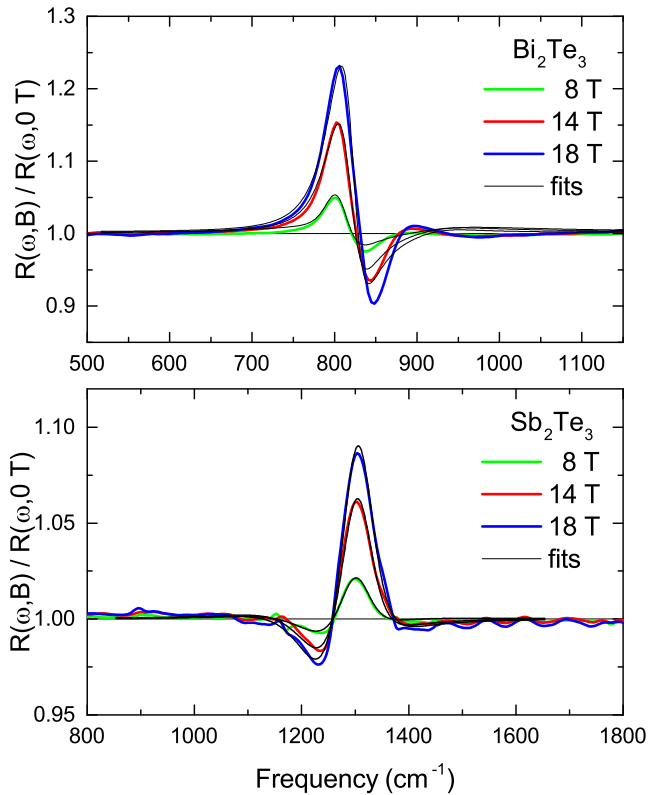}}%
\vspace*{-0.5cm}%
\caption{(Color online). (a) The best fits of reflectance
ratios of Bi$_2$Te$_3$ at several selected field values,
using Eq.~\ref{eq:dl-imag}. (b) The best fits of reflectance
ratios of Sb$_2$Te$_3$ at several selected field values,
using Eq.~\ref{eq:dlmag}.}
\vspace*{0.0cm}%
\label{fig:fits}
\end{figure}

In other to explore the origins of observed cyclotron antiresonance
we employ Drude model modified for the presence of magnetic
field \cite{lax67}. In spite of its simplicity, this model has been
very successful in
describing magneto-optical data in a variety of condensed
matter systems, most notably topological insulators
\cite{laforge10,kuzmenko14,dordevic12b,dordevic16}.
In the semiclassical approximation, the dynamics
of charge carriers (with effective mass {\it m})
in a solid is governed by \cite{marder-book}:

\begin{equation}
m \frac{d \vec{v} }{dt}+m \gamma \vec{v} = -e \vec{E} - e \vec{v} \times \vec{B_0}
\label{eq:lorentz}
\end{equation}
where the expression on the right is the Lorentz force, with constant magnetic field
B$_0$ being applied along z-axis. Assuming that both $\vec{v}$ and $\vec{E}$
vary as e$^{-i \omega t}$, one can solve \cite{marder-book} for the
complex dielectric function for the left and right
circularly polarized light $\tilde{\varepsilon}_{\pm}(\omega)$ as:

\begin{equation}
\tilde{\varepsilon}_{\pm}(\omega) = \varepsilon_{\infty} +
\frac{\omega_{p}^{2}}{ - \omega^2 - i \gamma \omega \mp
\omega_{c} \omega}
\label{eq:dlmag}
\end{equation}
where $\omega_p$ is the oscillator strength of the resonance mode
and $\gamma$ is its width. $\varepsilon_{\infty}$ is the high-frequency
dielectric constant. $\omega_c$ = eB$_0$/m is the cyclotron resonance frequency,
and is conventionally taken as positive for
electron-like and negative for hole-like carriers.
Using this model we could obtain good fits of reflectance
ratios of Sb$_2$Te$_3$. In Sb$_2$Te$_3$ (and Bi$_2$Te$_3$) charge carriers
are believed to be hole-like \cite{dordevic13}, so $\omega_c$ is
taken as negative. The best fits are shown in Fig.~\ref{fig:fits}(b) with
black lines. We notice that the model is capable of capturing all the important
features of Sb$_2$Te$_3$ data, in particular, the dip-peak structure
characteristic of cyclotron resonance seen in other systems, such
as Bi$_2$Se$_3$ \cite{dordevic16}.

On the other hand, we have not been able to obtain satisfactory
fits of reflectance ratios of Bi$_2$Te$_3$ for any values of model parameters.
In particular, the model cannot reproduce the peak-dip structure
characteristic of Bi$_2$Te$_3$ data. To circumvent this problem we propose the
idea that in Bi$_2$Te$_3$ cyclotron resonance frequency acquires imaginary
(or more generally complex) values. Imaginary cyclotron resonance frequency
is necessary in order to introduce an additional phase shift in the dielectric
function (Eq.~\ref{eq:dlmag}). The idea was inspired by the recently proposed
theory to explain quantum fluctuating superconductivity \cite{davison16}.
The authors used a complex cyclotron frequency, and referred to it as the
super-cyclotron resonance.

Cyclotron resonance is a direct consequence of applied magnetic field (Eq.~\ref{eq:lorentz})
and therefore in order to obtain imaginary cyclotron frequency, one must modify the
expression for the Lorentz force that magnetic field exerts on charge carriers.
We propose that in the case of Bi$_2$Te$_3$ the magnetic part of the Lorentz force
should be modified to:

\begin{equation}
\vec{F} = - e \frac{1}{\gamma} \frac{d \vec{v}}{dt} \times \vec{B_0}.
\label{eq:mod-lorentz}
\end{equation}
Other, more complicated modifications to the Lorentz force are also
possible.
Making the same assumptions as before \cite{marder-book}, and solving in
the semi-classical approximation as before (Eq.~\ref{eq:lorentz}), one
obtains for the complex dielectric function:

\begin{equation}
\tilde{\varepsilon}_{\pm}(\omega) = \varepsilon_{\infty} +
\frac{\omega_{p}^{2}}{ - \omega^2 - i \gamma \omega \mp
i \omega^2 \omega_{c}^{'} / \gamma}
\label{eq:dl-imag}
\end{equation}
which, compared with Eq.~\ref{eq:dlmag}, effectively has an imaginary
cyclotron resonance frequency $\omega_{c}^{'}$. We note however, that the
physical meaning of $\omega_{c}^{'}$ is different from $\omega_{c}$
(from Eq.~\ref{eq:dlmag}) and the two should not be directly compared.
We also point out that
Eq.~\ref{eq:dl-imag} satisfies the causality relations
$\tilde{\varepsilon}_{\pm}(-\omega)$ = $\tilde{\varepsilon}_{\mp}^{\ast}(\omega)$.

Using this model (Eq.~\ref{eq:dl-imag}) we were able to
obtain satisfactory fits for Bi$_2$Te$_3$.
The results of the fits are shown with black lines
in Fig.~\ref{fig:fits}, for several magnetic field values.
As can be seen, the fits can capture all the essential features
of the data, in particular the peak-dip structure of Bi$_2$Te$_3$.
We find this to be significant, as the conventional model (Eq.~\ref{eq:dlmag})
cannot fit the data for any values of fitting parameters.
It should also be noted that, even though the model Eq.~\ref{eq:dl-imag}
captures the most important feature of the data, it does not
reproduce the data around 900~cm$^{-1}$. This secondary, much weaker
structure might be due to Landau level transitions, and additional
terms might be needed in Eq.~\ref{eq:dl-imag} to account for it.


\begin{figure}[p]
\vspace*{-0.5cm}%
\centerline{\includegraphics[width=9.5cm]{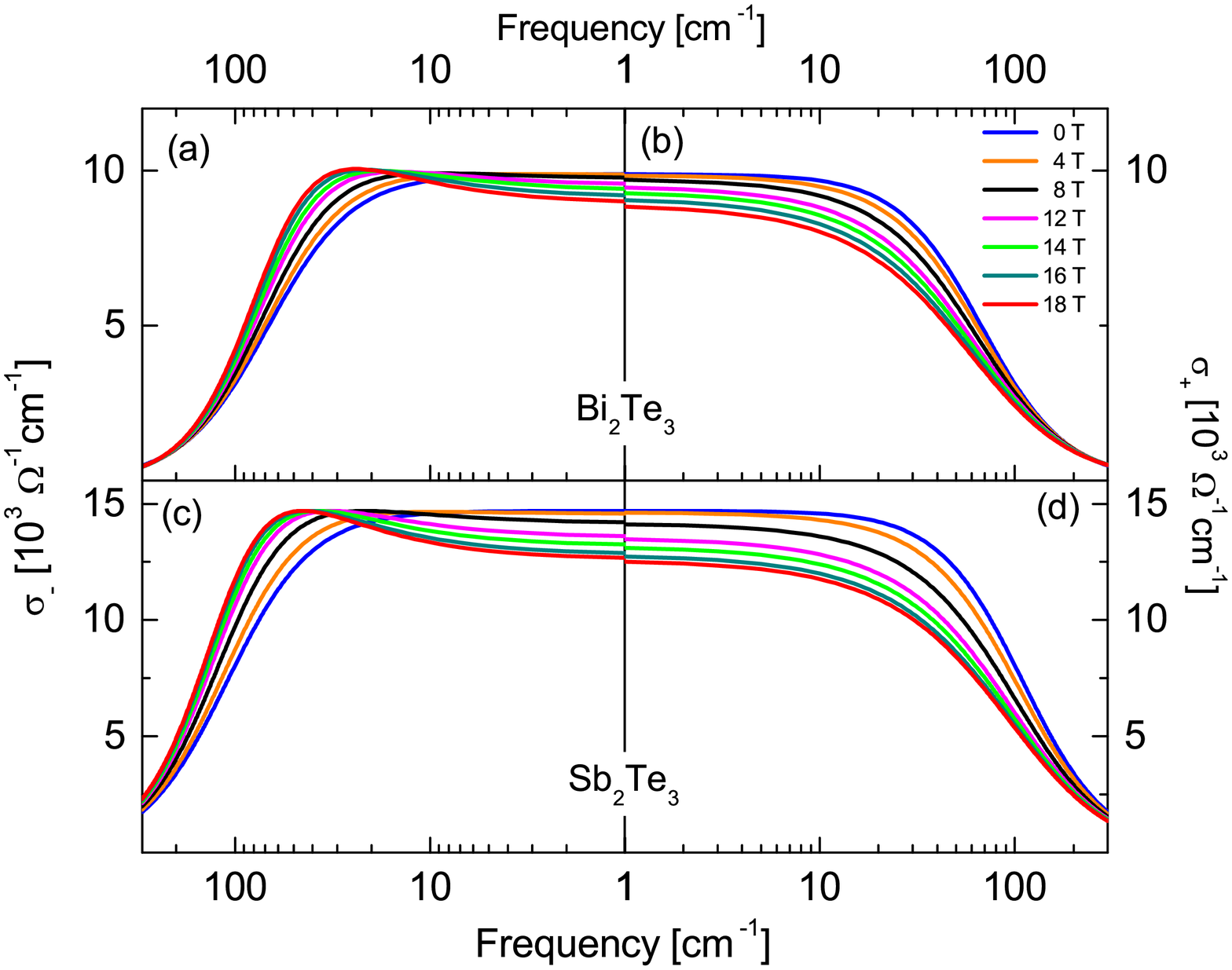}}%
\vspace*{0.0cm}%
\caption{(Color online). (a) and (b) Optical conductivities
for left and right circularly polarized light
$\sigma_{+}(\omega)$ (right panel) and $\sigma_{-}(\omega)$ (left panel)
for Bi$_2$Te$_3$ at different magnetic fields, generated from
the best fits of reflection ratios (Fig.~\ref{fig:fits}).
(c) and (d) Optical conductivities $\sigma_{+}(\omega)$ and
$\sigma_{-}(\omega)$ for Sb$_2$Te$_3$. The same
color coding is used in all four panels.}
\vspace*{-0.5cm}%
\label{fig:sigma}
\end{figure}

In order to get a better insight into carrier dynamics, we
explored other optical functions of both Bi$_2$Te$_3$ and Sb$_2$Te$_3$.
From the best fits of reflectance obtained with Eqs.~\ref{eq:dlmag}
and \ref{eq:dl-imag}
we generated optical conductivities for left and right
circular polarizations of light
$\sigma_{\pm}(\omega) =  \omega (\varepsilon_{\pm}(\omega) -1)/(4 \pi i)$.
Fig.~\ref{fig:sigma} displays the real part of
circular conductivity for both Bi$_2$Te$_3$ (top panels)
and Sb$_2$Te$_3$ (bottom panels).
For Sb$_2$Te$_3$ in zero field both $\sigma_{-}(\omega)$ and
$\sigma_{+}(\omega)$ are Drude modes. As the field increases,
$\sigma_{+}(\omega)$ is gradually suppressed, but maintains its
Drude shape. On the other hand, the peak in $\sigma_{-}(\omega)$
is shifted to finite frequencies, and is gradually suppressed,
but appears to maintain its width. At 18 T, the peak is at
approximately 45~cm$^{-1}$.
This behavior is typical of topological insulators, and
has been seen before, for example in Bi$_2$Se$_3$ \cite{armitage15,dordevic16}.

Surprisingly, the behavior of both $\sigma_{-}(\omega)$
and $\sigma_{+}(\omega)$ (Fig.~\ref{fig:sigma}(a) and (b) respectively)
in Bi$_2$Te$_3$ is similar to Sb$_2$Te$_3$, Bi$_2$Se$_3$
and other topological insulators. Even though the model used
to fit the data was different, all the features in the $\sigma_{\pm}(\omega)$
spectra are qualitatively similar. In zero field both $\sigma_{+}(\omega)$
and $\sigma_{-}(\omega)$
are Drude modes. As the field increases, $\sigma_{+}(\omega)$ is
suppressed, whereas in $\sigma_{-}(\omega)$
the peak gradually shifts to finite frequencies and at 18 T it
is at 24 cm$^{-1}$.


\begin{figure}[t]
\vspace*{0.0cm}%
\centerline{\includegraphics[width=9.5cm]{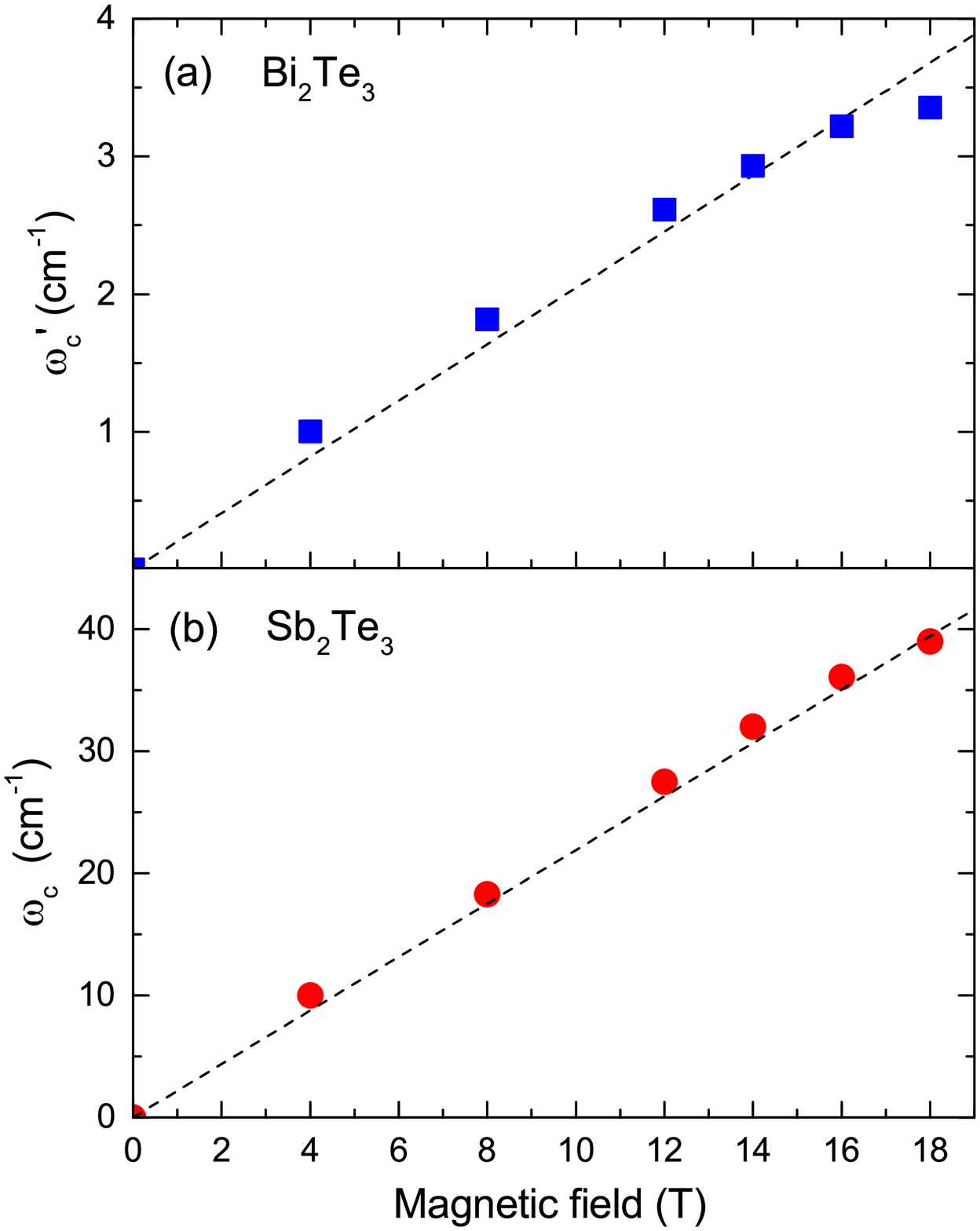}}%
\vspace*{-0.5cm}%
\caption{(Color online). (a) Magnetic field dependence of cyclotron frequency
$\omega_{c}^{'}$ in Bi$_2$Te$_3$. Dashed line is a
linear fit of the data. (b) Cyclotron resonance frequency $\omega_c$
of Sb$_2$Te$_3$. Dashed line is a linear fit,
from which the cyclotron effective mass of 0.42 m$_e$ was extracted.}
\vspace*{0.0cm}%
\label{fig:parameters}
\end{figure}

In Fig.~\ref{fig:parameters}
we plot the parameters of the best fits from Eqs.~\ref{eq:dlmag} and \ref{eq:dl-imag}.
Fig.~\ref{fig:parameters}(a) and (b) display the cyclotron resonance
frequencies $\omega_{c}^{'}$ and $\omega_c$ for Bi$_2$Te$_3$ and Sb$_2$Te$_3$ respectively.
In both compounds we observe linear field dependence which
extrapolates to zero in zero field, characteristic
of charge carriers with parabolic band dispersion.
Linear fits (shown with dashed lines) yield
$\hbar \omega_c$/B = 0.025 meV/T and 0.27 meV/T
for Bi$_2$Te$_3$ and Sb$_2$Te$_3$ respectively. In the case
of Sb$_2$Te$_3$  we can estimate the
cyclotron effective mass m/m$_e$~=~ 0.42. This
value is in general agreement with
earlier reports from quantum oscillations
measurements \cite{kohler76,kulbachinskii99}.


\section{Summary}

In summary, magneto-optical study of Bi$_2$Te$_3$ and Sb$_2$Te$_3$
has reveal a cyclotron resonance in the spectra of Sb$_2$Te$_3$,
and an antiresonance in Bi$_2$Te$_3$. To explain the behavior of
Bi$_2$Te$_3$ we introduced the idea of an alternative Lorentz force,
which resulted in an imaginary cyclotron frequency.
Physical interpretation of
this unconventional behavior requires deeper theoretical analysis
of how light couples to charge carriers in Bi$_2$Te$_3$.


The authors thank A.B. Kuzmenko, Y. Zhang and  X.R. Wang for useful
discussions. Work at Brookhaven is
supported by the US DOE under Contract No. DE-SC00112704 (H.L.
and C.P.). A portion of this work was performed at the National High
Magnetic Field Laboratory, which is supported by National Science
Foundation Cooperative Agreement No. DMR-1157490 and the State of Florida.

$^{\ast}$ Present address: Department of Physics, Renmin University,
Beijing 100872, China

$^{\dag}$ Present address: College of Physical Science and
Technology, Sichuan University, Chengdu, 610064, China

%
%

\begin{thebibliography}{10}

\bibitem{hartmann-book} H. Hartmann and K.-P. Wanczek, {\it Ion Cyclotron
Resonance Spectrometry}, Springer (1978).

\bibitem{guest-book} G. Guest, {\it Electron Cyclotron Heating of Plasmas},
Wiley-VCH, (2009).

\bibitem{lominadze-book} D.G. Lominadze, {\it Cyclotron Waves in Plasma},
Pergamon (2013).

\bibitem{cardona-book} P.Y. Yu and M. Cardona, {\it Fundamentals
of Semiconductors}, Springer (2010).

\bibitem{cr-book} S. Bhattacharya and K.P. Ghatak, {\it Effective Electron Mass
in Low-Dimensional Semiconductors}, Springer (2013).

\bibitem{kuzmenko11} I. Crassee, J. Levallois, A.L. Walter, M. Ostler,
A. Bostwick, E. Rotenberg, T. Seyller, D. van der Marel and A.B. Kuzmenko,
Nature Physics {\bf 7}, 48 (2011).  

\bibitem{armitage15} L. Wu, W-K. Tse, M. Brahlek, C.M. Morris,
    R.V Aguilar, N. Koirala, S. Oh and N.P. Armitage, Physical Review
    Letters {\bf 115}, 217602 (2015).  

\bibitem{dordevic16} S.V. Dordevic, G.M. Foster, M.S. Wolf, N. Stojilovic, H. Lei,
C. Petrovic, Z. Chen, Z.Q. Li and L.C. Tung, J.Phys.: Condens. Matter,
{\bf 28}, 165602 (2016). 

\bibitem{kuzmenko14} J. Levallois, P. Chudzinski, J.N. Hancock,
A.B. Kuzmenko and D. van der Marel, Phys.Rev.B {\bf 89}, 155123 (2014). 

\bibitem{kuzmenko16} P.J. de Visser, J. Levallois, M.K. Tran, J.-M. Poumirol,
I.O. Nedoliuk, J. Teyssier, C. Uher, D. van der Marel and A.B. Kuzmenko,
Phys.Rev.Lett. {\bf 117}, 017402 (2016).   

\bibitem{laforge12} A. Koncz, A.D. LaForge, Z Li, A. Frenzel, B. Pursley,
    T. Lin, X. Liu, J. Shi, S.V. Dordevic and D.N. Basov 2010
    {\it http://meetings.aps.org/link/BAPS.2010.MAR.H15.4}   

\bibitem{schafgans12} A.A. Schafgans, K.W. Post, A.A. Taskin, Y. Ando, X.-L. Qi,
B.C. Chapler and D.N. Basov, Phys.Rev.B {\bf 85}, 195440 (2012). 

\bibitem{dordevic12b} S.V. Dordevic, M.S. Wolf, N. Stojilovic, M.V. Nikolic,
    S.S. Vujatovic, P.M. Nikolic and L.C. Tung, Phys.Rev.B, {\bf 86},
    115119 (2012).

\bibitem{dordevic14} S.V. Dordevic, G.M. Foster, N. Stojilovic, E.A. Evans,
Z.G. Chen ,Z.Q. Li, M.V. Nikolic, Z.Z. Djuric, S.S. Vujatovic and
P.M. Nikolic, Phys. Status Solidi B {\bf 251}, 1510 (2014). 

\bibitem{shao17} Y. Shao, K.W. Post, J.-S. Wu, S. Dai, A.J. Frenzel,
A.R. Richardella, J.S. Lee, N. Samarth, M.M. Fogler, A.V. Balatsky,
D.E. Kharzeev and D.N. Basov, Nano Letters {\bf 17},  980 (2017).   

\bibitem{tung16} L.-C. Tung, W. Yu, P. Cadden-Zimansky, I. Miotkowski,
Y.P. Chen, D. Smirnov and Z. Jiang Physical Review B {\bf 93}, 085140 (2016).  

\bibitem{fisk89} Z. Fisk and J.P. Remeika, {\it Handbook on the
    Physics and Chemistry of Rare Earths}, edited by Gschneider K A
    and Eyring J (Amsterdam:Elsevier), Vol. 12 (1989).

\bibitem{canfield92} C.P. Canfield and Z. Fisk, Philosophical
    Magazine B {\bf 65}, 1117 (1992).

\bibitem{hunter98} B. Hunter, {\it Rietica - A visual Rietveld
    program}, International Union of Crystallography Commission on
    Powder Diffraction Newsletter No. 20, (Summer)
    http://www.rietica.org (1998).

\bibitem{dordevic13} S.V. Dordevic, M.S. Wolf, N. Stojilovic, H. Lei and C Petrovic,
    J.Phys.: Condens. Matter, {\bf 25},  075501 (2013). 

\bibitem{comm-flake} A cyclotron resonance was obsreved in a recent
    magneto-transmission study of Bi$_2$Te$_3$ nano-flakes \cite{tung16}.

\bibitem{lax67} B. Lax and J.G. Mavroides, {\it Semiconductors
    and Semimetals}, edited by R.K. Willardson and A.C. Beer,
    Academic Press, New York and London (1967).

\bibitem{laforge10} A.D. LaForge, A. Frenzel, B.C. Pursley, T. Lin,
    X. Liu, J. Shi and D.N. Basov, Physical Review B {\bf 81}, 125120 (2010).  

\bibitem{marder-book} M.P. Marder, {\it Condensed Matter Physics}, Wiley (2000).

\bibitem{kohler76} H. Kohler, Physica Status Solidi B {\bf 74}, 591 (1976).

\bibitem{kulbachinskii99} V.A. Kulbachinskii, N. Miura, H. Nakagawa,
    H. Arimoto, T. Ikaida, P. Lostak and C. Drasar, Phys.Rev.B
    {\bf 59}, 15733 (1999).    

\bibitem{davison16} R.A. Davison, L.V. Delacretaz, B. Gouteraux and S.A. Hartnoll,
    Phys.Rev.B {\bf 94}, 054502 (2016). 

\end{thebibliography}



\end{document}